\documentstyle[12pt]{article}
\begin{document}
\def\thefootnote{\fnsymbol{footnote}}
\begin{flushright}
KANAZAWA-95-04\\
April, 1995
\end{flushright}
\vspace*{13mm}
\begin{center}
{\Large\bf Chaotic Inflation based on an Abelian \\
D-flat Direction}
\vspace{1.5cm}\\
{\Large Daijiro Suematsu}
\footnote[1]{e-mail:suematsu@hep.s.kanazawa-u.ac.jp}\vspace{3mm}\\
{\Large and}\vspace{3mm}\\
{\Large Yoshio Yamagishi}
\footnote[2]{e-mail:yamagisi@hep.s.kanazawa-u.ac.jp}
\vspace{1cm}\\
{\it Department of Physics, Kanazawa University\\
 Kanazawa 920-11, JAPAN}
\vspace{1.5cm}
\end{center}
{\large\bf Abstract}\\
We study the inflation due to the D-flat direction of an extra $U(1)$.
This scenario is a hybrid of a right-handed sneutrino inflaton scenario
and a gauge non-singlet inflaton scenario.
The inflaton is a gauge non-singlet field which induces a right-handed
neutrino mass spontaneously through an extra $U(1)$ D-flat direction.
This right-handed neutrino mass can explain the solar neutrino
problem.
The reheating temperature resulting from the decay of the coherent
oscillation of the right-handed sneutrino is sufficiently high
so that the baryogenesis based on
the lepton number asymmetry can be applicable.
We also discuss the realistic model building.
\newpage
\setcounter{footnote}{0}
\def\thefootnote{\arabic{footnote}}
The inflation is now believed as a basic idea to resolve various
cosmological problems such as the flatness problem, the horizon problem
and so on\cite{kt,tu}.
It also seems to be supported by the observation of an anisotropy of the
cosmic microwave background by the Cosmic Background Explorer(COBE)
satellite\cite{wri}.
However, as is often referred, no satisfactory candidates of an inflaton
are not known within particle physics models still now.
They are usually introduced only for the inflation without any
substantial motivations from particle physics.
And they are considered to be very weakly interacting to have a flat
potential.
This brings various difficulties from the view points of particle
physics.

The basic features required for the inflaton may be summarized as :\\
(i)~to give a sufficiently long exponential expansion period for the
explanation of the flatness and the horizon problem and also
produce the sufficient entropy,\\
(ii)~to bring a suitable reheating of the universe to make the certain
baryogenesis scenario applicable after the inflation,\\
(iii)~to explain the microwave background anisotropy $(\delta T/T)_Q
\sim 6 \times 10^{-6}$, which has been recently detected by the COBE.

Recently some candidates of the inflaton have been proposed on the basis
of the motivations coming from particle physics.
Here we want to quote two interesting candidates.
One is the right-handed sneutrino scenario proposed by Murayama
{\it et al.}\cite{msyy}.
This is mainly motivated by the supersymmetry and also the explanation
of the small neutrino mass, which can resolve the solar neutrino
problem.
The other ones are gauge non-singlet fields which have a D-flat
direction of an extra $U(1)$\cite{ls}.
Such fields often appear in the models inspired by superstring, which
is now considered as the most promising unified theory
including gravity.
Although these have sufficient motivations and are also interesting
enough,
there seem to remain some unsatisfactory features.
The right-handed sneutrino scenario satisfies the above three
conditions.
In the model of ref.\cite{msyy}, however, the bare mass whose origin
is not known is introduced for the right-handed neutrino by hand.
This seems to be unsatisfactory from the view point of particle physics,
 where all masses of particles are usually considered to be
generated spontaneously.
On the other hand, the gauge non-singlet inflation scenario can not
satisfy the condition (ii)
unless we introduce a special kind of baryon number violating
interactions.
Although these usually exist in superstring-inspired models, they are
dangerous for the phenomena such as the proton decay.
This is caused by the fact that the gauge non-singlet inflaton is too
light to produce sufficiently high reheating temperature.

In this brief report we propose an inflation scenario in the
supersymmetric model.
It looks like a hybrid of these two scenarios and also
resolves their faults.
We study it in detail and also discuss the realistic model building.
In order to make our scenario definite, firstly we want to make some
basic assumptions and explain the features of our model.
We will return to these assumptions later
in relation to the model building.

A part of the superpotential relevant to the inflation is assumed to be
\cite{cve,sy}
\begin{equation}
W = \sum_{i,j} h^{ij}L_iHN_j + \sum_i {a_0 \over M_{\rm pl}}
\bar{\cal N}^2N^2_i
    + {a_n \over M_{\rm pl}^{2n-3}}({\cal N}\bar{\cal N})^n,
\end{equation}
where $M_{\rm pl}\sim 1.2\times 10^{19}$~GeV is the Planck mass and
$a_0$ and $a_n$ are constants, which will be determined in the following
analysis.
An integer $n$ is assumed to satisfy $n \ge 3$.
The chiral superfields $N_i$ and ${\cal N}$ have the similar
quantum number as a right-handed neutrino except for having a
non-trivial extra $U(1)$ charge $y_N$.
Because of its Yukawa coupling $h^{ij}L_iHN_j$ with the lepton
doublets $L_i$,
$N_j$ is regarded as a right-handed neutrino superfield.
The suffices $i$ and $j$ represent the generation.
And also $\bar{\cal N}$ is a chiral superfield which has an opposite
charge of the extra $U(1)$ against ${\cal N}$.\footnote{
For simplicity, we consider a model including only one pair of
$({\cal N}, \bar{\cal N})$.
We also do not consider the generation mixing of $N_j$ in the second
term of eq.(1).}
The structure of non-renormalizable parts of the superpotential is
similar to that of ref.\cite{ls}.
It is also assumed that there are no other higher
order terms for $N_i, {\cal N}$ and $\bar{\cal N}$ in eq.(1).
This is a very non-trivial assumption and we shall come back to this point
later.

The scalar potential for these fields is derived from eq.(1) as
\begin{eqnarray}
V&=&\sum_i{4a_0^2 \over M_{\rm pl}^2}\vert\tilde{\bar{\cal N}}\vert^4
\vert\tilde N_i\vert^2
+\left({na_n \over M_{\rm pl}^{2n-3}}\right)^2\vert\tilde{\cal N}
|^{2n-2}\vert\tilde{\bar{\cal N}}|^{2n}
+\left| \sum_i{2a_0 \over M_{\rm pl}}\tilde{\bar{\cal N}}\tilde N^2_i
+{na_n \over M_{\rm pl}^{2n-3}}
\tilde{\cal N}^n\tilde{\bar{\cal N}} ^{n-1} \right|^2
\nonumber \\
&+&\sum_i m_N^2\vert \tilde N_i\vert^2- m_{\cal N}^2(\vert \tilde{\cal N}
\vert^2
+\vert \tilde{\bar{\cal N}}\vert^2 ) \nonumber \\
&+&{1 \over 2}g^2y_N^2(\sum_i \vert \tilde N_i\vert^2
+\vert\tilde{\cal N}\vert^2-\vert \tilde{\bar{\cal N}}\vert^2)^2,
\end{eqnarray}
where $\tilde N_i$, $\tilde{\cal N}$ and $\tilde{\bar{\cal N}}$ represent
scalar components of each superfield, respectively.
The scalar masses in the second line are assumed to appear as a result of
the supersymmetry breaking.
The last line represents the D-term contribution of an extra $U(1)$.
As easily seen from eq.(2), this scalar potential is D-flat along the
direction of
$\displaystyle\sum_i \vert \tilde N_i\vert^2 +\vert\tilde{\cal N}\vert^2=
\vert\tilde{\bar{\cal N}}\vert^2$.
The right-handed neutrino $N_i$ can have the supersymmetric large mass
because of this D-flatness.
In fact, this scalar potential has a nontrivial minimum at
\begin{equation}
\vert\tilde N_i \vert =0, \qquad \vert \tilde{\cal N}\vert=
\vert\tilde{\bar{\cal N}}\vert
=\left[{m_{\cal N}^2M_{\rm pl}^{2(2n-3)} \over
(2n-1)n^2a_n^2}\right]^{1 \over 4n-4}.
\end{equation}
Because of these large vacuum expectation values of
$\tilde {\cal N}$ and $\tilde{ \bar{\cal N}}$,
the lepton number violation occurs\footnote{
There appears no massless Majoron associated to this lepton number
violation because it is eaten by the extra $U(1)$ gauge boson. }
and the mass of a right-handed neutrino is produced spontaneously
through the second term of $W$ as\cite{cve,sy}
\begin{equation}
M_N=a_0\left[{1 \over (2n-1)n^2a_n^2}\right]^{1 \over 2n-2}
(M_{\rm pl}^{n-2}m_{\cal N})^{1 \over n-1}.
\end{equation}
This relation constrains the coefficients $a_0$ and $a_n$ as
\begin{equation}
a_0a_n^{-1 \over n-1}=\left[(2n-1)n^2\right]^{1 \over 2n-2}
{M_N \over (M_{\rm pl}^{n-2}m_{\cal N})^{1 \over n-1}}. \nonumber
\end{equation}
If we require $M_N \sim 10^{11}$~GeV, the neutrino can get the Majorana
mass about $\sim 10^{-3}$~eV via the seesaw mechanism for a Dirac mass
$\sim 1$~GeV\cite{see}.
This is suitable for the MSW solution of the solar neutrino
problem\cite{solar}.
Taking $n=3$ and the soft scalar mass as $m_{\cal N}\sim 100$~GeV,
from eq.(5) we get
\begin{equation}
a_0\sim 7.5a_3^{1/2}.
\end{equation}
If $\tilde{\cal N}$ and $\tilde{\bar{\cal N}}$ quickly damp to the value
in eq.(3) along the D-flat direction, our model is expected to
reduce to that of ref.\cite{msyy} after their damping.

Now we consider the chaotic inflation of this model.
At the Planck epoch there is not enough time to realize the
thermal equilibrium and then every field is expected to be
out of equilibrium.
Some of them can have larger values than $M_{\rm pl}$ under the condition
that each term in the Lagrangian is
${^<_\sim}~O(M_{\rm pl}^4)$\cite{chao}.
In the present model, if the coefficients $a_0$ and $a_n$ are extremely
small, $\tilde N_i$ and $\tilde{\cal N}$ can have the large field values
 larger than $O(M_{\rm pl})$.
This is because these have a D-flat direction
and a renormalizable interaction of $N_i$ in the superpotential is only a
Yukawa coupling $h^{ij}L_iHN_j$,
whose effect is sufficiently small as discussed in ref.\cite{msyy}.
Among the scalar components $\tilde N_i$, $\tilde N_1$ which has the smallest
Yukawa coupling can have the largest field value.
Then we only consider its time evolution among $\tilde N_i$ hereafter.
In the following study we confine ourselves on the $n=3$ case.
And $\tilde N_1$, $\tilde{\cal N}$ and $\tilde{\bar{\cal N}}$ are
assumed to evolve along the
D-flat direction $\vert\tilde N_1\vert^2+\vert\tilde{\cal N}\vert^2
=\vert\tilde{\bar{\cal N}}\vert^2$.

For the successful inflation scenario our model should satisfy the
above mentioned three conditions (i)$\sim$(iii).
We also need to require the condition (6) to make the MSW
mechanism applicable to the solar neutrino problem.
These requirements constrain parameters $a_0$ and $a_n$ in the
superpotential.
If we put $\vert\tilde N_1\vert =u$ and $\vert\tilde{\cal N}\vert =v$,
the scalar potential along the D-flat direction becomes
\begin{eqnarray}
V&=&V_{a_0}+V_{a_3}+m_N^2u^2 -m_{\cal N }^2(u^2+2v^2),\nonumber \\
V_{a_0}&\equiv&4{a_0^2 \over M_{\rm pl}^2}(2u^6+3u^4v^2+u^2v^4),
\nonumber \\
V_{a_3}&\equiv&9{a_3^2 \over M_{\rm pl}^6}v^4(u^2+v^2)^2
(u^2+2v^2)+12{a_0a_3 \over M_{\rm pl}^4}
u^2v^3(u^2+v^2)^{3\over 2}.
\end{eqnarray}
As noticed above, the coefficient $a_0$ and $a_3$ are required to be
extremely small in the successful chaotic inflation.
And we find from eq.(6) that $a_0 \gg a_3$ should be satisfied.
Thus $V_{a_0}$ is expected to be a dominant part
of the scalar potential $V$ during the inflation era.
In the followings we shall determine $a_0$ and $a_3$ under
such a situation .

At first we consider the condition (i).
The time evolution equations of the field $\phi$ are
\begin{equation}
\ddot\phi + 3H\dot\phi +\Gamma_\phi\dot\phi=
-{\partial V \over \partial \phi} \qquad (\phi =u,~~ v),
\end{equation}
where $H$ is the Hubble parameter which is now approximated as
$H=\sqrt{8\pi V_{a_0} \over 3 M_{\rm pl}^2}$.
During this period the effective masses
$M_\phi^2\equiv ({\partial V \over \partial \phi})/\phi$
are estimated as
\begin{equation}
M_u^2 \sim {8a_0^2 \over M_{\rm pl}^2}(6u^4+6u^2v^2+v^4), \qquad
M_v^2 \sim {8a_0^2 \over M_{\rm pl}^2}(3u^4+2u^2v^2).
\end{equation}
{}From the form of $V_{a_0}$ we can take $u ~{^>_\sim}~ v>M_{\rm pl}$
as the initial values for these fields.
When $u~{^>_\sim}~ v > M_{\rm pl}$, both $u$ and $v$ slowly damp.
Once $H \sim M_u$ is realized, $u$ critically damps with a small
e-folding and begins to oscillate with a sufficiently large amplitude.
At this time $u$ becomes smaller than $M_{\rm pl}$ but $v$ still
remains larger than $M_{\rm pl}$.
This results in $M_v<H<M_u$ and causes the above phenomenon.
After $u$ starting the oscillation, $v$ continues the slow
rolling.\footnote{
The condition $a_0 \gg a_3$ coming from the right-handed
neutrino mass makes $\tilde{\cal N}$ inflaton generally.
In the present scenario a right-handed sneutrino $\tilde N_i$
seems to be
difficult to play a role of inflaton.}
When $H \sim M_v$, $v$ begins to oscillate and damps to its vacuum
value rapidly.\footnote{
This comes from the features of $H$ and $M_v$. It is recently
used to solve the cosmological moduli problem\cite{drt}.}
Through this era, $u$ continues oscillating without any effective
damping.\footnote{These picture has been checked by the
numerical calculation.}
After $v$ conversing its vacuum value, our model reduces to the one of
ref.\cite{msyy}.

The validity of the slow-roll approximation for $v$ after $u$ starting
the oscillation is justified under the conditions\cite{st}
\begin{equation}
\epsilon(v)=\left({M_{\rm pl}^2 \over 16\pi}\right)
\left({V^\prime_{a_0}(v) \over V_{a_0}(v)}\right)^2 < 1, \qquad
\eta(v)={M_{\rm pl}^2 \over 8\pi}
{V^{\prime\prime}_{a_0}(v) \over V_{a_0}(v)}< 1.
\end{equation}
where $V_{a_0}(v)\sim {4a_0^2 \over M_{\rm pl}^2}u_{\rm av}^2v^4$
with the averaged value of $u^2$.\footnote{
To estimate the averaged value of $u^2$ we used the numerical analysis
in the self-consistent way.
In the following arguments we use that result
$u_{\rm av} \sim 10^{-2}M_{\rm pl}$.}
These parameters can be estimated as
\begin{equation}
\epsilon(v)\sim {2 \over 3}\eta(v), \qquad
\eta(v)\sim {3 \over 2\pi}\left({M_{\rm pl} \over v}\right)^2.
\end{equation}
Therefore, the slow-roll approximation is valid as far as
$v_{\rm end}~ {^>_\sim}~0.7M_{\rm pl}$.
Eq.(8) is reduced to $3H\dot v =-V^\prime_{a_0}$
in a region where the above slow roll condition is satisfied.
Thus the e-folding of the expansion
between $v_{\rm in}$ and $v_{\rm end}$ can be
expressed as
\begin{equation}
N(v_{\rm in}, v_{\rm end})=-\int^{v_{\rm end}}_{v_{\rm in}}
{8\pi \over M_{\rm pl}^2}{V_{a_0} \over V^\prime_{a_0}}dv \sim
{\pi \over M_{\rm pl}^2}(v_{\rm in}^2-v_{\rm end}^2).
\end{equation}
The condition for the sufficient inflation is given as
$N(v_{\rm in}, v_{\rm end})~{^>_\sim}~60$.
This requires $v_{\rm in}~ {^>_\sim}~4.4M_{\rm pl}$
when $v_{\rm end}~ {^>_\sim}~0.7M_{\rm pl}$.

Next we impose the condition (iii).
Following the usual argument on the scalar density perturbation,
the microwave background quadrupole anisotropy $(\delta T/T)_Q$ is
expressed as
\begin{equation}
\left| \left({\delta T \over T}\right)_Q\right| \sim
\sqrt{\left. {32\pi \over 45}{V^3_{a_0} \over V^{\prime 2}_{a_0}M_{\rm pl}^6}
\right|_{k \sim H}}
={2 \over 3}\sqrt{2\pi \over 5}{a_0u_{\rm av} \over M_{\rm pl}}
\left({v \over M_{\rm pl}}\right)^3_{k \sim H}
\sim 0.13\left({a_0u_{\rm av} \over M_{\rm pl}}\right)N_H^{3 \over 2}
\end{equation}
where $N_H\equiv \pi({v \over M_{\rm pl}})^2_{k\sim H}$.
This value is estimated when the scale $k^{-1}$, which corresponds to
the present horizon size, crossed inside the horizon during
inflation\cite{tu}.
If we put $N_H \sim 50$ to realize $(\delta T/T)_Q\sim 6\times 10^{-6}$,
we find that we should take $a_0 \sim 1.3\times 10^{-5}$.
Assuming this value for $a_0$, the requirement eq.(6) from the
right-handed neutrino mass determines $a_3$ as
$a_3=2.9 \times 10^{-12}$.
For these values the above estimated condition of the sufficient
inflation is trivially satisfied.
{}From eq.(3), for these parameters the true minimum is realized at
$\vert \tilde{\cal N}\vert =\vert\tilde{ \bar{\cal N}}\vert
\sim 10^{-1.6}M_{\rm pl}$.
As explained above, $v$ quickly damps to its vacuum values
after the end of inflation.
When $v$ reaches to its true vacuum, $u$ still continues
oscillating
with the sufficiently large amplitude of order $10^{-2}M_{\rm pl}$
and dominates the energy density.
This coherent oscillation of $u$ decays to the light particles
through the Yukawa couplings satisfying the D-flat condition.
Thus our model is expected to be equivalent to one of ref.\cite{msyy}
after the inflation.

The reheating temperature
$T_{RH} \sim 0.1\sqrt{M_{\rm pl}\Gamma_{N_1}}$ is crucial
to consider what kind of baryogenesis scenario can work.
$\Gamma_{N_1}$ is the decay ratio of the field $\tilde N_1$.
In the present model the reheating is expected to occur due to the
decay of the oscillation of $\tilde N_1$ to the light particles
as ref.\cite{msyy}.
In this process, if CP violation exists, the lepton number asymmetry
is produced.
This asymmetry will be converted to the baryon number asymmetry
through the electroweak anomalous process if the reheating temperature
is high enough.
As seen in the previous part, $\tilde {\cal N}$ reaches its true vacuum
 soon after the end of the inflation because of its critical damping.
Therefore $\tilde{\cal N}$ is irrelevant to
the production of the lepton number asymmetry, which occurs at the later
stage.
The decay of $\tilde N_1$ is mediated by the Yukawa coupling
$h^{i1}L_iH \tilde N_1$ and the decay ratio is estimated as
$\Gamma_{N_1} \sim {\vert h^{13}\vert^2 \over 4\pi}M_N $.
Here $h^{13}$ is the largest Yukawa coupling of
$\tilde N_1$ to the charged
lepton and we take it as $h^{13} \sim 10^{-4}$ to $10^{-5}$.
The right-handed neutrino mass $M_N$ is $\sim 10^{11}$~GeV
so that the reheating temperature can be estimated as
$T_{RH} \sim 10^8$~GeV to $10^9$~GeV.
This reheating temperature seems to be suitable for the baryogenesis
based on the lepton number asymmetry\cite{fy}.
In fact, following the detailed discussion of ref.\cite{msyy}
the expected maximum baryon number to entropy ratio $Y_B(\equiv n_B/s)$
is estimated as\cite{ksht}
$$Y_B = {8n_g+4 \over 22n_g +13}Y_L^{\rm max} $$
by using
the generated maximum lepton number to entropy ratio $Y_L^{\rm max}$
and the generation number $n_g=3$.
If we take $\epsilon$ as the asymmetry in the decay of $\tilde N_1$
into leptons and antileptons, $Y_L^{\rm max}$ is expressed as
$Y_L^{\rm max} \sim \epsilon {3 \over 4}{T_{RH} \over M_N}$.
In order to realize the correct baryon number asymmetry
$Y_B \sim 10^{-10}$, we need $\epsilon \sim 10^{-7}$ to $10^{-8}$
for the above mentioned reheating temperature.
This value of $\epsilon$ is somehow smaller than the one of
ref.\cite{msyy} mainly coming from our smaller setting of $M_N$.
To make $\epsilon$ larger we can consider the dilution effect discussed
in ref.\cite{msyy}.

Here we summarize the features of our model again.
In the present model
the field $\tilde {\cal N}$ plays the role of inflaton due to its
slow-roll property as ref.\cite{ls} and produces the right-handed
neutrino mass spontaneously through its abelian D-flat direction.
The reheating occurs by the decay of $\tilde N_1$ with the large mass,
 which can explain the solar neutrino problem.
We can get the sufficiently high temperature which is
convenient for the baryogenesis.
In the present model the gravitino mass is considered as $O(1)$~TeV as
usual and the reheating temperature is $T_{RH} \sim 10^8$~GeV
to $10^9$~GeV.
These values make our model free from the gravitino problem\cite{gra}.

Finally we discuss the construction of particle physics models
which have the features discussed in this brief report.
It is very interesting that our model seems to be naturally
embedded in the certain class of superstring-inspired models.
There often appear the extra $U(1)$ symmetries accompanied with
the singlet fields as $N_i$, ${\cal N}$ and $\bar{\cal N}$.
Such a concrete model is discussed in ref.\cite{sy}.
In the above arguments we require the particular type of the
superpotential $W$ and also the extremely small coefficients $a_0$
and $a_3$.
As is well-known, in the superstring theory there are a lot of
discrete symmetries
which can constrain the form of the superpotential $W$.
We can expect that it may at least explain the non-existence of the
lower order terms in the non-renormalizable ones in $W$ due to
such discrete symmetries\cite{disc}.
Moreover, it is shown that in a certain type of superstring the
non-renormalizable terms are produced only through the non-perturbative
world sheet instanton effects.
There appears an exponentially small suppression factor
$\exp(-c/g^2)$, where $c>0$ and $g$ is a world sheet coupling
constant\cite{nonp}.
This may explain the reason of the smallness of coefficients $a_0$ and $a_3$
as suggested in ref.\cite{ls}.
These facts are favorable for the present scenario to be realized in realistic
particle physics models.
However, there remains a problem on supergravity corrections to
the scalar potential.
This problem has been seriously discussed in the inflation scenario
within the supergravity framework\cite{sgra}.
In the present stage we can not say anything about this problem and
also why the higher order non-renormalizable terms do not appear in $W$.
A special form of the K\"ahler potential derived from superstring and
stringy symmetry may be relevant to such problems as suggested
in ref.\cite{cllsw}.
Anyway, this is a crucial problem when we consider the inflation
within the superstring framework.
\vspace{1cm}

We would like to thank Prof.~T.~Yanagida for his critical comments and
interests in the present work.
The work of D.~S. is partially supported in part by a Grant-in-Aid for
Scientific
Research from the Ministry of Education, Science and Culture(\#05640337
and \#06220201).
\newpage


\begin{thebibliography}{99}
\bibitem{kt}For a review, see ,for example,\\
E.~D.~Kolb and M.~S.~Turner, {\it The Early Universe}
(Addison Wesley, 1990),

A.~D.~Linde, {\it Particle Physics and Inflationary Cosmology}
(Harwood, Chur, Switzerland,1990).

\bibitem{tu}M.~S.~Turner, "Inflation after COBE",
FERMILAB-Conf-92/313-A and references therein.

\bibitem{wri}E.~L.~Wright, {\it et al}., Astrophys. J., {\bf 420}(1994)1.

\bibitem{msyy}H.~Murayama, H.~Suzuki, T.~Yanagida and J.~Yokoyama,
Phys. Rev. Lett. {\bf 70}(1993)1912.

\bibitem{ls}G.~Lazarides and Q.~Shafi, Phys Lett. {\bf B308}(1993)17.

\bibitem{cve}M.~Cveti$\check{\rm c}$ and P.~Langacker, Phys. Rev.
{\bf D46}(1992)R2759,

N.~Haba, C.~Hattori, M.~Matsuda, T.~Matsuoka and D.~Mochinaga,
Prog. Theor. Phys. {\bf 92}(1994)153.

\bibitem{sy}D.~Suematsu and Y.~Yamagishi, preprint KANAZAWA-94-22
(hep-ph/9411239).

\bibitem{see}M.~Gell-Mann, P.~Ramond and S.~Slansky, in Supergravity,
eds. P.~van Nieuwenhuizen and D.~Freedman (North-Holland, Amsterdam,
1979)p315.

T.~Yanagida, in Proc. Workshop on Unified Theory and Baryon Number in
the Universe, eds. A.~Sawada and H.~Sugawara(KEK, Tsukuba, Japan, 1979).

\bibitem{solar}S.~P.~Mikheyev and A.~Yu.~Smirnov, Yad. Fiz.
{\bf 42}(1985)[Sov. J. Nucl. Phys. {\bf 42}(1985)913.]

L.~Wolfenstein, Phys. Rev. {\bf D20}(1979)2634.

\bibitem{chao}A.~D.~Linde, Phys. Lett {\bf 129B}(1983)177.

\bibitem{drt}L.~Randall and S.~Thomas, preprint MIT-CTP-2331,
SCIPP94-16,

G.~Dvali, preprint IFUP-TH 09/95,

M.~Dine, L.~Randall and S.~Thomas, preprint SCIPP 95-13.

\bibitem{st}P.~J.~Steinhardt and M.~S.~Turner, Phys. Rev. {\bf D29}
(1984)2162.

\bibitem{fy}M.~Fukugita and T.~Yanagida, Phys. Lett. {\bf B174}(1986)45.

\bibitem{ksht}S.~Yu~Khlebnikov and M.~E.~Shaposhnikov, Nucl. Phys.
{\bf B308}885,

J.~S.~Harvey and M.~S.~Turner, Phys. Rev. {\bf D42}(1990)3344.

\bibitem{gra}S.~Weinberg, Phys. Rev. Lett. {\bf 48}(1982)1303,

G.~D.~Coughlan, W.~Fischler, E.~D.~Kolb, S.~Raby and G.~G.~Ross,
Phys. Lett. {\bf 131B}(1983)59.

\bibitem{disc}C.~A.~L${\rm\ddot{u}}$tkin and G.~G.~Ross, Phys. Lett.
{\bf B214}(1988)357.

P.~Zoglin, Phys. Lett. {\bf B228}(1988)47.

C.~Hattori, M.~Matsuda, T.~Matsuoka and H.~Mino, Prog. Theor. Phys.
{\bf 82}(1989)599.

\bibitem{nonp}M.~Cveti$\check{\rm c}$, Phys. Rev. {\bf 37}(1988)2366,

E.~Witten, Nucl. Phys. {\bf B268}(1986)79.

\bibitem{sgra}R.~Holman, P.~Ramond and G.~G.~Ross, Phys. Lett.
{\bf B137}(1984)343.

B.~A.~Ovrut and P.J.Steinhardt, Phys. Rev. {\bf D30}(1984)2061,

A.~B.~Goncharov and A.~D.~Linde, Phys. Lett. {\bf B139}(1984)27,

H.~Muratama, H.~Suzuki T.~Yanagida and J.~Yokoyama,
preprint YITP-U-93-29(1993),

K.~Kumekawa T.~Moroi and T.~Yanagida, Prog. Theor. Phys.
{\bf 92}(1994)437.

\bibitem{cllsw}E.~J.~Copeland, A.~R.~Riddle, D.~H.~Lyth, E.~D.~Stewart
and D.~Wands, Phys. Rev. {\bf D49}(1994)6410.

\end{thebibliography}
\end{document}